\documentclass[none, preprint, colorlinks, urlcolor=blue, citecolor=blue ]{imsart}\usepackage[]{graphicx}\usepackage[]{color}
\makeatletter
\def\maxwidth{ %
  \ifdim\Gin@nat@width>\linewidth
    \linewidth
  \else
    \Gin@nat@width
  \fi
}
\makeatother

\definecolor{fgcolor}{rgb}{0.345, 0.345, 0.345}

\usepackage{framed}
\makeatletter
 {\par\unskip\endMakeFramed%
 \at@end@of@kframe}
\makeatother

\definecolor{shadecolor}{rgb}{.97, .97, .97}
\definecolor{messagecolor}{rgb}{0, 0, 0}
\definecolor{warningcolor}{rgb}{1, 0, 1}
\definecolor{errorcolor}{rgb}{1, 0, 0}
\newenvironment{knitrout}{}{} 

\usepackage{alltt}
\usepackage{natbib}
\usepackage{float}
\usepackage{amsthm,amsmath}
\usepackage{hyperref}

\IfFileExists{upquote.sty}{\usepackage{upquote}}{}
\begin{document}

\begin{frontmatter}

\title{A note on sum and difference of correlated chi-squared variables}
\runtitle{On sum and difference of correlated chi-squared}

\begin{aug}
\author{\fnms{Alberto} \snm{Ferrari}\corref{}
\ead[label=ed1]{aferrari34@yahoo.com}}
\address{FROM research foundation\\ Ospedale Papa Giovanni XIII\\ Piazza OMS, 1, 24127\\ Bergamo (BG), Italy\\ \printead{ed1}}
\affiliation{FROM research foundation \thanksmark{t2}}
\thankstext{t2}{The views and opinions expressed in this article are those of the author and do not necessarily reflect the position of the institution.}

\runauthor{A. Ferrari}
\end{aug}

\begin{abstract}
Approximate distributions for sum and difference of linearly correlated \(\chi^{2}\) distributed random variables are derived. It is shown that they can be reduced to conveniently parametrized gamma and Variance-Gamma distributions, respectively.  

The proposed distributions are very flexible, and the one for sum in particular has straight-forward generalizations to cases where multiple \(\chi^{2}\) variables with different parameters are involved. The results promptly extend to every sum of gamma variables with common scale and to every difference between gamma variables with common shape and scale.

The fit of the distributions is tested on simulated data with remarkable results.The approximations presented are expected to be especially useful to researchers working on gamma-distributed variables.  

\end{abstract}



\end{frontmatter}


\section{Introduction}
\addcontentsline{toc}{section}{Introduction}
The distribution of the sum of two independent \(\chi^{2}\) distributed random variables with  \( m_{1} \)  and  \( m_{2} \)  degrees of freedom is known to be \(\chi^{2}\) with  \( m_{1}+m_{2} \)  degrees of freedom.

However, the case of non-independent variables is less straight forward. \cite{Gunst1973} derived the distribution of a sum of two linearly correlated \(\chi^{2}\) random variables with  \(  \rho <1 \). More recently,  \cite{Joarder2013} expressed the density for a sum of correlated \(\chi^{2}\) with common  \( m \) degrees of freedom in terms of a generalized hypergeometric function.

Here it is shown that the distribution of the sum of linearly correlated \(\chi^{2}\) variables with product moment correlation coefficient \( 0 \leq  \rho  \leq 1  \) can be approximated by a conveniently parametrized gamma distribution. By a similar method, the distribution of the difference can be derived as well in particular cases.

Both results have straight-forward generalizations to sums of gamma variables with common scale and to differences of identically distributed gamma variables. 

\section{Distribution of sum of correlated \(\chi^{2}\) variables}

Let 

 \[ X_{1} \sim  \chi ^{2} \left( m_{1} \right) , \] 

 \[ X_{2} \sim  \chi ^{2} \left( m_{2} \right) , \]

 \[ X=X_{1}+X_{2} .\]

We will first consider the case of  \( m = m_{1} = m_{2} \) . In the independent case, with shape and scale parametrization.

 \[ X \sim  \Gamma  \left( m, 2 \right).  \] 

Let us consider the case of the distribution of  \( X_{1} \)  multiplied by a constant. In this case, a well-known result is that

 \[ cX_{1} \sim  \Gamma  \left( \frac{m}{2},2c \right),  \] 

When  \( c=2 \) 

 \[ 2X_{1} \sim  \Gamma  \left( \frac{m}{2},4 \right).  \] 

When  \( X_{1}=X_{2} \),\   \( X=2X_{1} \) . Whereas previous derivations of the distribution of linear combinations of correlated \(\chi^{2}\) variable focus on  \(  \rho <1 \), here a different approach is used, akin to the method used by \cite{kotz1963distribution} to derive an approximate distribution for the sum of correlated gamma variables. These two results can be seen as special cases of the distribution of the sum of two positively correlated \(\chi^{2}\) variables where  \(  \rho  \left( X_{1},X_{2} \right) =0 \)  and \(   \rho  \left( X_{1},X_{2} \right) =1 \), respectively. 

Under this perspective, it can be inferred that the general distribution of X in the correlated case reduces to a gamma at the extremes, and therefore should be gamma or generalized gamma for different values of  \( 0 \leq  \rho  \leq 1 \). We can now deduce some other properties of this distribution based on considerations on expected value and variance.

Clearly:\par

 \[ E \left( X \right) =E \left( X_{1} \right) +E \left( X_{2} \right) =2m .\] \par

The expected value should not depend on correlation. Also, it follows from basic properties of variance that, in the correlated case: \par

 \[ Var \left( X \right) =Var \left( X_{1} \right) +Var \left( X_{2} \right) +2Cov \left( X_{1},X_{2} \right) ;\]
 
 since  \( Var \left( X_{1} \right) =Var \left( X_{2} \right) =2m ,\) \par

 \[ Cov \left( X_{1},X_{2} \right) = \rho \; \sqrt[]{Var \left( X_{1} \right) Var \left( X_{2} \right) }=2 \rho m ,\] \par

 \[ Var \left( X \right) =4m \left( 1+ \rho  \right)  .\] \par

The distribution of X should therefore be gamma or generalized gamma with expected value equal to the sum of the expected values, i.e.  \( 2m \), and variance equal to the one of the individual \(\chi^{2}\) scaled by a  \( 2 \left( 1+ \rho  \right)  \)  factor. There is no straight-forward way to parametrize the generalized gamma distribution so that these conditions are met, since it only reduces to a gamma when the second scale parameter is 1. However, it is very easy to achieve the result by choosing appropriate values for a simple two parameters gamma distribution. Since for the gamma distribution \(\mu=k\theta\) and \(\sigma^{2} = k\theta^{2}\), the distribution for the sum of two correlated \(\chi^{2}\) variables with common  \( m \)  should be approximately:\par

 \[ X \sim  \Gamma  \left( \frac{m}{1+ \rho },2 \left( 1+ \rho  \right)  \right)  .\] \par

The result is easily generalized to the case where  \( m_{1}  \neq  m_{2} \) . In this case: \par

 \[ E \left( X \right) =E \left( X_{1} \right) +E \left( X_{2} \right) =m_{1}+m_{2} ,\] \par

 \[ Cov \left( X_{1},X_{2} \right) =2 \rho \; \sqrt[]{m_{1}m_{2}} ,\] \par

 \[ Var \left( X \right) =2 \left( m_{1}+m_{2}+2 \rho  \; \sqrt[]{m_{1}m_{2}} \right)  ,\] \par

 \[ X \sim  \Gamma  \left( \frac{ \left( m_{1}+m_{2} \right) }{2 \left( 1+\frac{2 \rho  \; \sqrt[]{m_{1}m_{2}}}{m_{1}+m_{2}} \right) },2 \left( 1+\frac{2 \rho  \; \sqrt[]{m_{1}m_{2}}}{m_{1}+m_{2}} \right)  \right)  .\] \par

It is easily verified that the conditions concerning mean and variance are satisfied. \par
In the case of the sum of \(N \) variables with pairwise correlation coefficients \(\rho_{i,j}: i,j= 1, ..., N \),

 \[  \sum _{i=1}^{N}X_{i} \sim  \Gamma  \left( \frac{ \sum _{i=1}^{N}m_{i}}{u},u \right)  ,\] 
 
 \[ u=2 \left( 1+\frac{2 \sum _{i \neq j}^{N} \rho _{i,j} \; \sqrt[]{m_{i}m_{j}}}{ \sum _{1=1}^{N}m_{i}} \right)  .\] \par
 
\begin{figure}[H]

\begin{knitrout}
\definecolor{shadecolor}{rgb}{0.969, 0.969, 0.969}\color{fgcolor}
\includegraphics[width=\maxwidth]{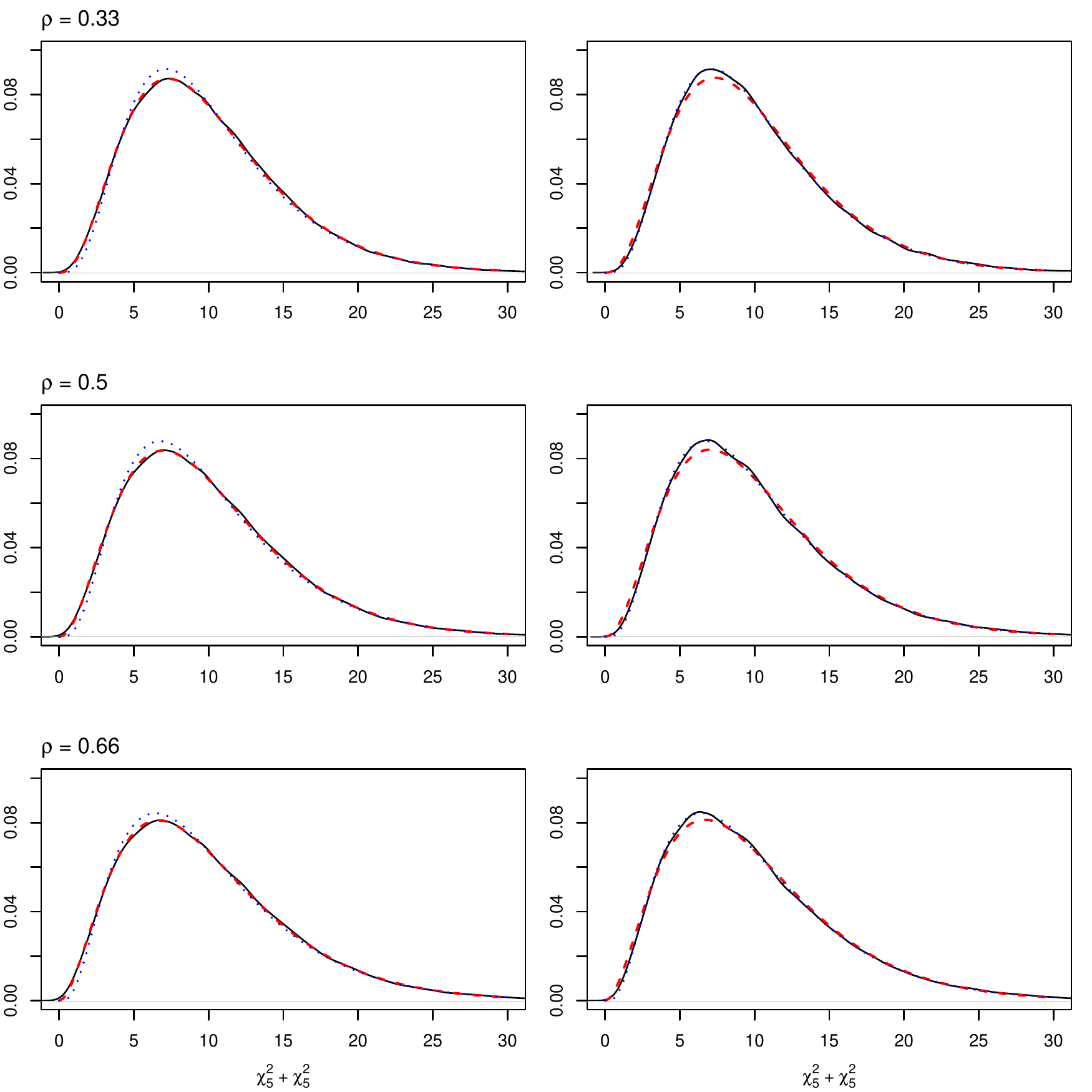} 

\end{knitrout}
\caption{Empirical distribution and density function for a sum of \(\chi^{2}\) variables with m = 5. Variables on the left are simulated by the quantile function method, whereas variables on the right are drawn from a bivariate \(\chi^{2}\) distribution. Solid line: empirical distribution; dashed red: gamma distribution; dotted blue: exact distribution.\label{fig:myrcode1}}
\end{figure}

 Figure \ref{fig:myrcode1} shows empirical distribution in the case of  \( m=m_{1}=m_{2}=5 \), with overlapping density function of the gamma model and the exact distribution derived by Joarder et al. Data were simulated both as random draws from a bivariate  \(\chi^{2}\) and by the quantile function method, i.e. by transformation of a bivariate normal variate; the results are compared between the two methods.

Interestingly, whereas, as expected, the gamma approximation has a slightly worse fit than the exact distribution on data simulated from a bivariate \(\chi^{2}\), on the other hand it shows a better fit when non-independence has a different functional form.

\begin{figure}[H]

\begin{knitrout}
\definecolor{shadecolor}{rgb}{0.969, 0.969, 0.969}\color{fgcolor}
\includegraphics[width=\maxwidth]{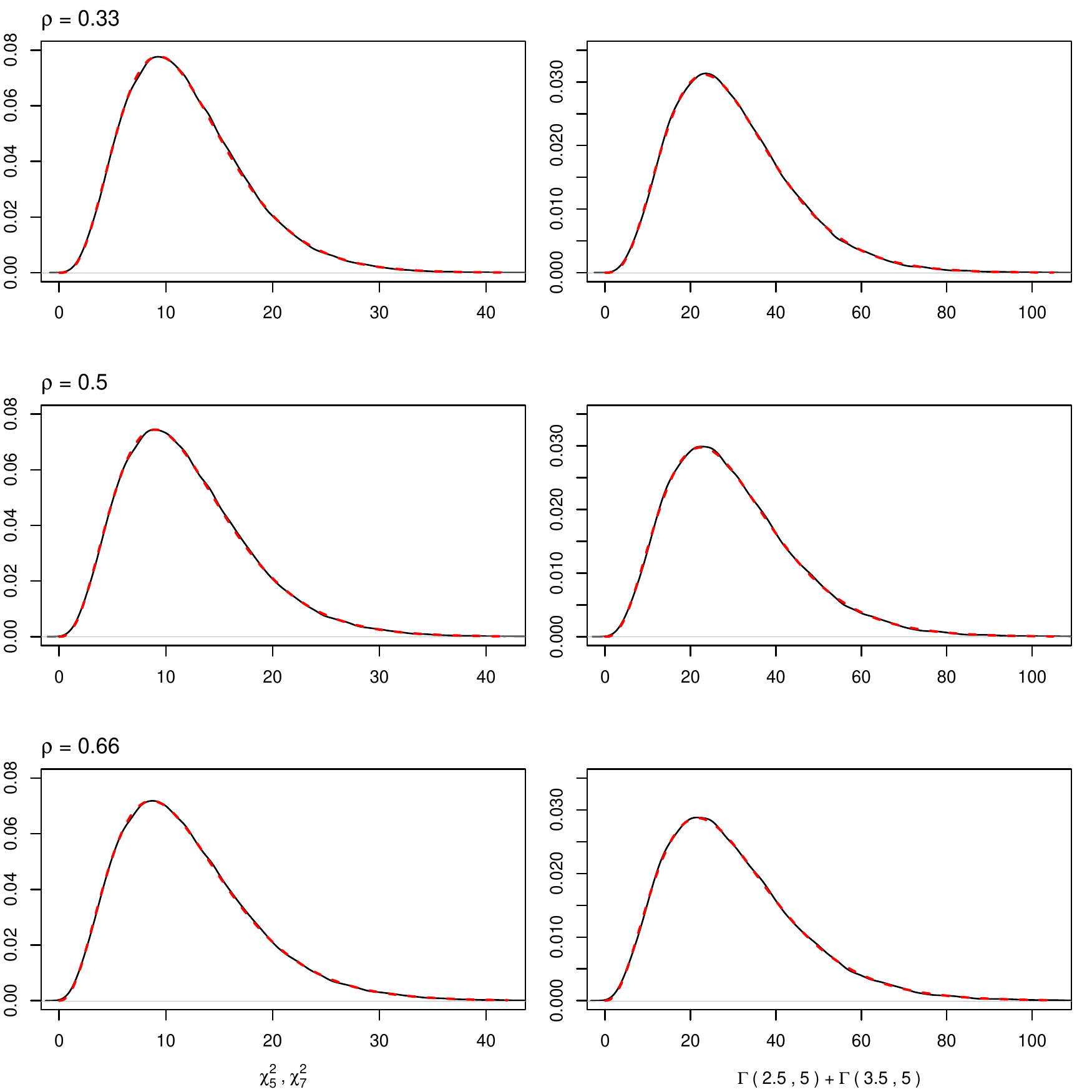} 

\end{knitrout}
\caption{Empirical distribution (solid black) with overlapping gamma density function (dashed red) for a sum of \(\chi^{2}\) variables with \(m_{1}=5\)\:and\:\(m_{2}=7 \) and for a sum of correlated gamma distributions with common \(\theta\).\label{fig:myrcode2}}
\end{figure}

 Figure \ref{fig:myrcode2} shows empirical distribution and overlapping density function for some combinations of parameters and correlation coefficients; data in this case were simulated by the quantile function method. 
 
Since the \(\chi^{2}\) is just a gamma distribution with shape \(k=\frac{m}{2}\) and scale \(\theta = 2\), the approach can also be extended to any sum of correlated gamma variables with common scale parameter \(\theta\). If \(X_{1}, ... ,X_{N}\) are gamma variables with shape parameters \(k_{1}, ... ,k_{N}\), pairwise correlation coefficients \(\rho_{i,j}: i,j= 1, ..., N \) and common \(\theta\) then 
 
  \[  \sum _{i=1}^{N}X_{i} \sim  \Gamma  \left( \frac{ \sum _{i=1}^{N}k_{i}}{u},\theta u \right)  ,\] \par
 
 with 
 
 \[ u=1+\frac{2 \sum _{i \neq j}^{N} \rho _{i,j} \; \sqrt[]{k_{i}k_{j}}}{ \sum _{1=1}^{N}k_{i}} .\] \par
 
Figure \ref{fig:myrcode2} also displays the distribution fit for the sum of two correlated gamma distributions. 

\section{Distribution of difference of correlated \(\chi^{2}\) variables}
A similar line of reasoning allows to approximate the distribution of the difference of two correlated \(\chi^{2}\) variables, at least in the case where  \( m = m_{1} = m_{2} \) . \par

Let \par

 \[ X=X_{1}-X_{2} .\] \par

Now when \(X_{1}\) and \(X_{2}\) are indipendent \(X\) is Variance-Gamma distributed [\cite{klar2015note}]. In particular we see from the moment generating function of the \(\chi^{2}\) distribution that\par

 \[ M_{X_{1}} \left( t \right) =M_{X_{2}} \left( t \right) = \left( 1-2t \right) ^{\frac{-m}{2}} ,\] \par

 \[ M_{X} \left( t \right) =M_{X_{1}} \left( t \right) M_{X_{2}} \left( -t \right) = \left( 1-4t^{2} \right) ^{\frac{-m}{2}}= \left( \frac{\frac{1}{4}}{\frac{1}{4}-t^{2}} \right) ^{\frac{m}{2}} ,\] \par

which is the moment generating function of the Variance-Gamma distribution, under generalized hyperbolic parametrization:

 \[ M_{X} \left( t \right) =e^{ \mu t} \left( \frac{ \alpha ^{2}- \beta ^{2}}{ \alpha ^{2}- \left(  \beta ^{2}+t^{2} \right) } \right) ^{ \lambda } ,\] \par

with  \(  \mu =0,  \alpha =\frac{1}{2},  \beta =0,  \lambda =\frac{m}{2} \) .\par

We change to the parametrization proposed by \cite{Seneta2004} for convenience, which in this case leads us to:\par

\[ c= \mu =0, \] 

\[  \sigma =\sqrt[]{\frac{2 \lambda }{ \alpha ^{2}- \beta ^{2}}}=\frac{\sqrt[]{2 \lambda }}{ \alpha }=2  \; \sqrt[]{m}, \] 

 \[  \nu =\frac{1}{ \lambda }=\frac{2}{m}, \] 

\[  \theta = \beta =0. \] 

We can use the same approach as above and look at this distribution as a special case of difference of correlated \(\chi^{2}\) variables with  \(  \rho =0 \) . However, the approach cannot be faithfully replicated, since in the case of  \(  \rho =1 \)  the distribution is degenerate with zero mean and variance. Here it is only known that the distribution becomes Variance-Gamma in the special case of  \(  \rho =0 \) . However, we know that in the more general case the distribution must have zero mean and that its variance must tend to zero as  \(  \rho  \)  approaches 1. \par

The mean of the Variance-Gamma distribution is\par

 \[ E \left( X \right) =c+ \theta  .\] \par

With \( c= \theta =0 \) ,  \( E \left( X \right) =0 \) . \par

The variance of the Variance-Gamma distribution is:\par

 \[ Var \left( X \right) = \sigma ^{2}+ \theta ^{2} \nu , \] \par

which here becomes \par

 \[ Var \left( X \right) = \sigma ^{2}=4m .\] \par

From properties of variance we also know that, in the correlated case:\par

 \[ Var \left( X \right) =4m \left( 1- \rho  \right) , \] \par

i.e. the variance should be scaled by  \( 1- \rho  \)  factor to account for correlation. This is accomplished by multiplying  \(  \sigma  \)  by  \( \sqrt[]{1- \rho } \) .\par

The distribution of a difference between two correlated \(\chi^{2}\) variables is therefore expected to be approximately\par

 \[ X \sim VarianceGamma \left( 0,2\;\sqrt[]{m \left( 1- \rho  \right) },0,\frac{2}{m} \right) . \] 

for  \( 0 \leq  \rho <1 \) . As  \(  \rho  \)  approaches 1 the variance approaches 0. 

The generalization to a difference of two gamma distributions with common \(k\) and \(\theta\) is very straight-forward and is obtained by setting \(\nu = \frac{1}{k}\) and \(\sigma=\theta \; \sqrt{2k}\).
If \(X_{1}\) and \(X_{2}\) are gamma variables with common shape and scale \(k, \theta\), then 
 \[ X_{1}-X_{2} \sim VarianceGamma \left( 0,\theta\;\sqrt[]{2 k \; \left( 1- \rho  \right) },0,\frac{1}{k} \right) . \]\par

\begin{figure}[H]

\begin{knitrout}
\definecolor{shadecolor}{rgb}{0.969, 0.969, 0.969}\color{fgcolor}
\includegraphics[width=\maxwidth]{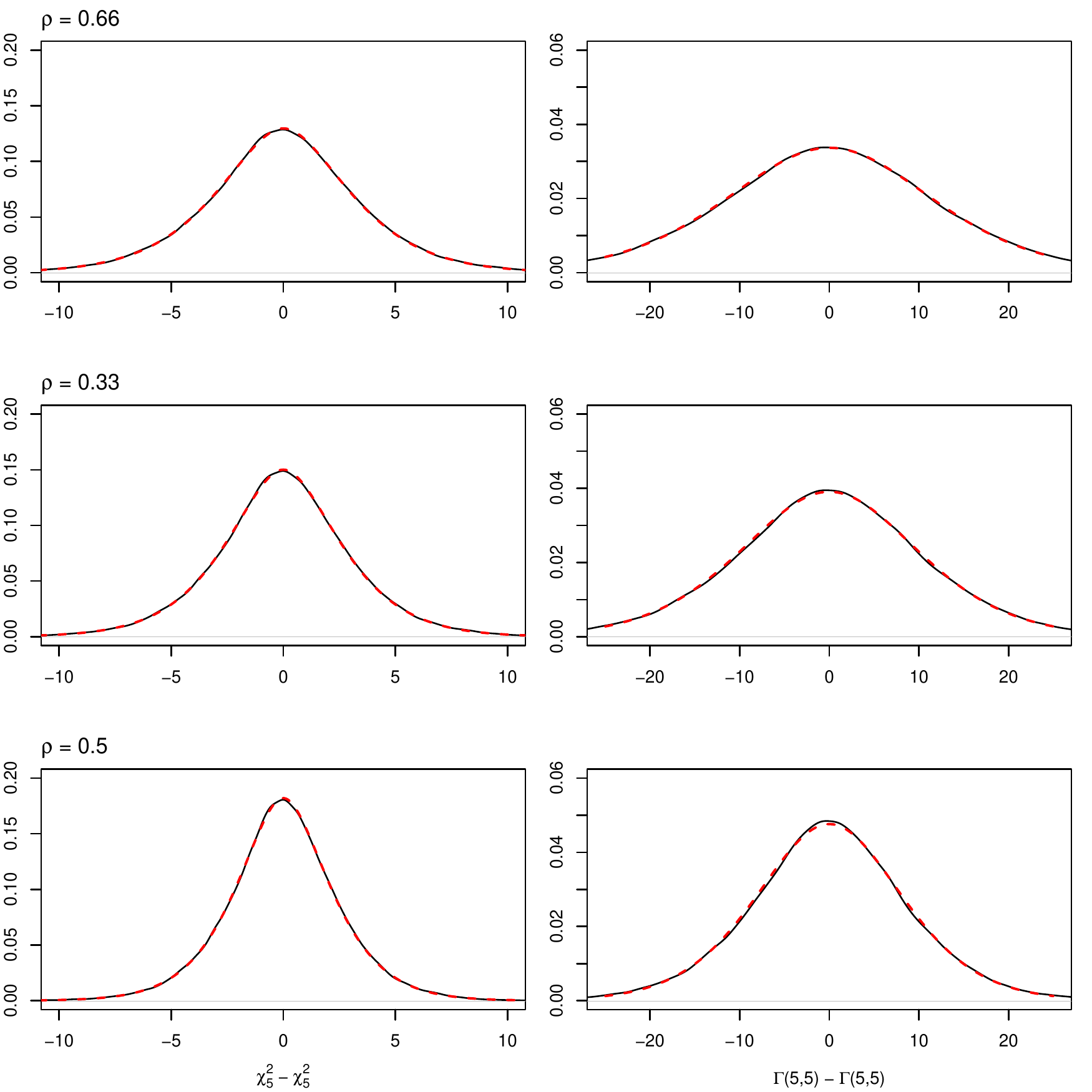} 

\end{knitrout}
\caption{Empirical distribution (solid black) with overlapping Variance-Gamma density function (dashed red) for the difference of two \(\chi^{2}\) variables drawn from a bivariate \(\chi^{2}\) distribution with m = 5 and for the difference of two arbitrary gamma distributions. \label{fig:myrcode3}}
\end{figure}

The fit of the distribution is tested on 100.000 simulations; Empirical distribution and overlapping density function are shown in Figure \ref{fig:myrcode3} for different values of \(\rho\).\par
 
\section{Conclusions}
In this work it is shown that the distribution of the sum of a series of correlated \(\chi^{2}\) variables can be approximated by an adequately parametrized gamma distribution. The exact density function for said distribution has been derived before; however the approximation that is proposed here is much simpler and has more general applicability, since it generalizes easily to cases where multiple variables with different degrees of freedom are summed. As shown in simulations, the distribution closely fits the data; remarkably, whereas the fit was slightly worse than the exact distribution in data simulated from squaring multivariate normal variates, it was actually better when data were simulated using the quantile function method. This is likely due to the approximation being less reliant on specific distributional assumptions, which makes the method attractive to those who work with correlated gamma variables when little is known about the functional form of non-independence. \par

In the second part of the work it is shown, in a similar fashion, that the distribution of a difference between two correlated \(\chi^{2}\) variables with common degrees of freedom is, at least approximately, a variance gamma distribution with 0 mean and  \(  \sigma  \)  equal to  \( 2m \;\sqrt[]{1- \rho } \).  The density function fits the data in simulations. \par
Both results are generalized to sum and difference of gamma variables with common scale paramaters. These approximations are attractively simple, which given their very good fit to the empirical distributions suggest they could have widespread applications in statistics.

\renewcommand{\refname}{\vspace*{-1em}}
\specialsection*{References}
 
\bibliography{citation}

\end{document}